# Knowledge linkage structures in communication studies using citation analysis among communication journals

Han Woo PARK[a] and Loet LEYDESDORFF[b]

**Abstract**

This research analyzes a "who cites whom" matrix in terms of aggregated, journal-journal citations to determine the location of communication studies on the academic spectrum. Using the *Journal of Communication* as the seed journal, the 2006 data in the *Journal Citation Reports* are used to map communication studies. The results show that social and experimental psychology journals are the most frequently used sources of information in this field. In addition, several journals devoted to the use and effects of media and advertising are weakly integrated into the larger communication research community, whereas communication studies are dominated by American journals.

[a] Department of Communication & Information, YeungNam University 214-1, Dae-dong, Gyeongsan-si, Gyeongsangbuk-do, South Korea, Zip Code 712-0749; parkhanwoo@hotmail.com http://www.hanpark.net
[b] Amsterdam School of Communications Research (ASCoR), University of Amsterdam, Kloveniersburgwal 48, 1012 CX Amsterdam, The Netherlands; loet@leydesdorff.net, http://www.leydesdorff.net

## Introduction

Given the notable institutional build-up of media and communication studies since 1990, in not only North America but also other regions including East Asia and Western Europe, the number of manuscripts published in the Institute for Scientific Information (ISI)-rated journals is increasing rapidly (the ISI has recently been renamed Thomson-Reuters but the name ISI continues to be widely used). The conventional perception of communication studies is that it is an incomplete aggregation of atomized research domains.

Scholars have viewed communication studies as a heterogeneous group with different attributes (Delia, 1987; Leydesdorff, 2004), partly due to the duality of communication studies. As Rogers (1994, p. 48) stressed, "Communication is a professional field, as well as a scientific discipline. The mass media industries stand behind the academic field of communication, offering jobs for its graduates, helping fund its research, and providing endowments for professors and schools of communication." As a result, communication studies have split into many academic entities; for example, speech communication, mass communication, advertising, and journalism. Continual developments in media technology have also forced the founding of new specialties that include telecommunication, new media, computer-mediated communication, informatics, and human-computer interaction.

In 2005, communication studies gained legitimacy as an independent discipline for educating and studying the nature of human communication mediated by various channels (Treadwell, 2006). The National Science Foundation of the USA (NSF) formally accredited communication studies as an academic field with a particular subject in the NSF's list of disciplines. The representative professional International Communication Association (ICA) was founded more than half a century ago in 1950. Derek de Solla Price (1963) noted that big science refers to a field that has established theories, mature research methods, and scientific collaborations. In this regard, communication studies are ready to become accepted as a big science in the 2000s.

The success in having communication studies recognized as a stand-alone discipline inspired us to write this article. We address the following two research questions: Where on the academic spectrum do communication studies reside? Which subject category is core to communication studies? In order to investigate these topics, we focus on linkage structures in communication studies using citation analysis among communication journals. Citations do not occur in a social vacuum. What communication researchers cite forms how communication studies are constituted. More specifically, we evaluate the development and progress of communication studies by collecting a journal-journal citation matrix from the ISI's database as an input file to our scholarly investigation.

**Related studies**

Compared to the speedy diffusion of communication and media studies, scientometric exercises that examine the academic structure of current communication research have been rare. In response to this, the *Journal of Communication*, a top journal published by the largest association of scholars in communication studies, the ICA, has published two studies that analyze the national origins of communication journals (Lauf, 2005) and professional careers of article authors (Bunz, 2005). Both articles look at the position of communication studies as an independent discipline using bibliometric data obtained from scholarly journals and their authors. In a similar vein, Masip (2005) focused on research articles written by European authors in thirty-five communication journals.

There are more restricted studies that focus on reviewing some specific area in communication studies: Internet research related to advertising (Cho & Khang, 2006), research trends in mass communication (Kamhawi & Weaver, 2003), and media research on the Internet (Kim & Weaver, 2002).

More recently, the Communication Research Centre (CRC, 2007) at the University of Helsinki conducted a comprehensive project about current trends of media and communication research in seven countries (Finland, USA, Germany, France, Japan, Estonia, and Australia). According to the reports about individual countries, communication studies around the world are somewhat differently deployed in some major strands. While in the U.S.-based academic tradition, quantitative investigations of media effects are dominant, communication programs with different approaches (e.g., humanistic and sociological practices) are gradually growing. The CRC's report on the U.S. also provides the top ten communication and media journals using the ISI's impact factor from 1998 to 2005. At the top of the top ten list (according to an eight-year mean) is *Public Opinion Quarterly*, followed by *Communication Research*, and the *Journal of Communication*. *Media Psychology, Discourse & Society, Human Communication Research, Cyberpsychology & Behavior, Public Culture, Political Communication*, and the *Journal of Health Communication* are also included in the top-ten citation list.

These previous studies are limited to individual characteristics of communication studies. The past literature is unable to discover the networked traits within communication studies by ignoring structural properties among communication journals and their neighboring journals in other fields. As alluded earlier, communication studies as a disciplinary field are made of several scientific domains. Perhaps, this field can be captured better through network-oriented indicators and visualizations than by indicators which normalize over the journals subsumed under the ISI-category of “communication.” The connectivity patterns represented in the citation behavior of authors in communication studies journals may reveal how this fragmented community perceives and reconstructs its relevant scholarly environments.

**Methods: Social network metrics and data collection**

Indicators and visualizations based on social network analysis are used in this research (for a detailed explanation about social network metrics and technical procedures, see Hanneman & Riddle, 2005). Previous studies (Kim et al., 2006; Leydesdorff, 2005; Park et al., , 2005; Park & Thelwall, 2006; Park & Leydesdorff, 2008) have shown that the structural pattern of citations among a set of authors, articles, or journals can be better examined using a network perspective. For instance, Freeman (2004) suggested that a network approach utilizing citation data may be a more robust means of studying the inside world of an invisible research community. Using social network techniques, Freeman (2004, pp. 165-166) showed that there were surprisingly few bridging linkages between physicists and the traditional social network analysts who cited Milgram (1967)’s famous article entitled “The Small World Problem”. The network perspective, developed within the social and behavior sciences, is currently identified as an influential method in library and information science, and scientometrics (Otte & Rousseau, 2002; Thelwall, 2004, pp. 213-217).

Among the many techniques for gathering network data, snowballing is frequently used in the case where the entire number of nodes in a given social system and whether they entertain relations are *a priori* unclear (Garton et al., 1997). For example, researchers often get a particular person to list their acquaintances and report their interactions with all other network members in terms of face-to-face meeting frequency per week, the type of communication media used, and the closeness of relationship. Leydesdorff (2007) took advantage of this network sampling technique to visually map the specific journal's relevant citation environment and succeeded in detecting a collection of journals that make a network having a particular journal as an entrance point. More recently, Park and Leydesdorff (2008) applied this sampling technique to a scientometrics exercise by focusing on the cited dimension of a Korean ISI-listed journal.

The data for this research were harvested from the CD-Rom versions of the Journal Citation Reports (JCR) 2006. More specifically, the 107 journals in the 2006 ISI database that cited the *Journal of Communication* were taken and an asymmetrical (valued) matrix—that is, a matrix in which the frequencies of citations between journals are provided—was constructed for social network analysis. Next, ISI's subject categories and publication places of individual journals listed in the sample were collected. Additionally, the journals cited in the *Journal of Communication* were gathered and a citation matrix among these (154) journals was made. However, our analysis focuses on the domain of 107 journals in the citation impact environment. Aggregated citation data and journal names for these journals are provided in Appendix I. All values in the JCR-data refer to unique citation relations at the article level.

The *Journal of Communication* was selected as the focal journal for the mapping of communication studies because, as signaled in its title, this Journal is a flagship publication in the communications research community. The ICA, the world's largest scholarly body in terms of the participatory countries, has an institutional responsibility for the Journal. Compared to the ICA's other titles (e.g., *Communication Theory, Human Communication Research,* and *Journal of Computer-Mediated Communication*), the Journal is relatively open to many studies with varied theoretical and methodological backgrounds. Most articles published in *Human Communication Research,* for example, seem to contain a purely quantitative analysis. Therefore, studying the journal structure of communication studies as a discipline with the *Journal of Communication* as an exemplar is probably the best approach, although this is indeed a specialist universe of publications.

UciNet for Windows (Borgatti et al., 2002), a commonly used software program for network analysis, was used to calculate the various metrics and NetDraw packaged with UciNet was used for the graphic illustrations. In social network research, visualization is often an informative tool to reveal the relationship between different network nodes, in this case, journals. NetMiner software (Cyram, 2003) was also used for the visualizations.

## Results

### *Citation structure among 107 journals citing the Journal of Communication*

The *Journal of Communication* was cited by 107 journals in 2006. The network analysis of the 107 * 107 citation matrix reveals that the clustering coefficient of this network is 0.59. The clustering coefficient—defined as the proportion of links between the vertices within the neighbourhood of a vertex divided by the number of links that could possibly exist between all these vertices—indicates the degree to which friends of a person know each other (for a mathematical definition, see Watts, 1999). In this research, the clustering coefficient measures the extent to which a journal's neighboring journals are connected to one another. Averaging these proportions over all journals in the network shows how closely interconnected neighbors in the network are. Interestingly, the coefficient value (0.59) found in the study is very close to the result (0.56) that network physicists have identified in social networks of scientific collaboration (Barabási, 2002).

The diameter value of 5.00 implies that all journals are connected to the extent of five hops when the direction of the citations is not considered. The mean distance is 2.08. That is, if two journals in the network are selected they are on average about two steps away from each other. Reciprocity is 0.55: more than half of all the journals exchange their citations. For example, the authors of journal A chose the articles published in journal B as their references and so did the journal B's authors with reference to journal A. All of these network metrics (of course, with the exception of reciprocity) were done after binarization and symmetrization.

More specifically, the *Psychological Bulletin* is the journal most cited by others (1,524 times) in this network. Conversely, the *Journal of Communication* was cited 781 times. In network analysis, "indegree centrality" refers to the total number of citations a journal receives from the other journals at the article level and "outdegree centrality" measures how many times a journal cites the articles published in other journals in the same network (Freeman, 1979). For the calculations of both these centrality measures, within-journal self citations were excluded.

Degree centrality measures are primary indicators for the selected node's network activity. The dominance of the *Psychological Bulletin* in terms of the being-cited dimension, that is, indegree centrality, means that research articles in communication science journals connected to the *Journal of Communication* provide references mainly to psychology and its related concepts. As Delia (1987, p. 23) stated, the most influential sources for the construction of communication research come from theoretical concepts and methodological behaviorism in psychology, and the quantitative research orientation in sociology and political science, that is, those specialties among the social sciences in which issues of measurement and operational procedures are of central concern.

In the citing dimension, *Psychological Bulletin* had a relatively small number of citations which make reference to 267 articles published in 23 other journals in the network. The most central journal in the citing dimension is *Sex Roles* with 813 citations to 46 other journals, followed by *Social Science & Medicine*, the *Journal of Youth & Adolescence*, and the *Journal of Communication* with 635, 542, and 512 citations, respectively. The difference between these highest numbers in the cited and citing dimensions are reflected in the network centralization metrics. The outdegree network centralization (citing dimension) is 32.73 percent while the indegree centralization (cited dimension) is 85.11 percent. The higher this percentage the more centralized a network is. Therefore, these 107 journals are connected to each other in the cited network more than in the citing network. In the dominant dimension (cited), the *Psychological Bulletin* is the most central source of citations.

Table 1 lists the most highly and the most rarely cited ten journals with their respective citation counts. Figure 1 illustrates the 107 journals according to their indegrees, that is, times cited in the network. The most highly cited *Psychological Bulletin* occupies the central position of the target board and the next most frequently cited journals are assorted up and down. The remaining journals are scattered around the target.

**Table 1**. The most highly and most rarely cited journals in a network using the *Journal of Communication* as the seed journal

| Highly cited ten journals | Indegree (times cited) | Least cited ten journals | Indegree (times cited) |
|---|---|---|---|
| *PsycholBull* | 1524 | *NewZealJPsychol* | 9 |
| *DevPsychol* | 848 | *BritJEducTechnol* | 7 |
| *JCommun* | 781 | *InteractLearnEnvir* | 5 |
| *CommunRes* | 674 | *WorldEcon* | 4 |
| *SocSciMed* | 623 | *Policing* | 3 |
| *SexRoles* | 552 | *ZKlPsychPsychoth* | 2 |
| *PsycholRep* | 388 | *FoodDrugLawJ* | 0 |
| *HumCommunRes* | 386 | *LangLearnTechnol* | 0 |
| *PsycholWomenQuart* | 362 | *PolitSci* | 0 |
| *JBroadcastElectron* | 351 | *TextTalk* | 0 |

- The abbreviations used for the journal titles are available at http://images.isiknowledge.com/help/WOS/A_abrvjt.html. (See also Appendix I.)

**Figure 1**. Indegree distributions of the 107 journals under study.

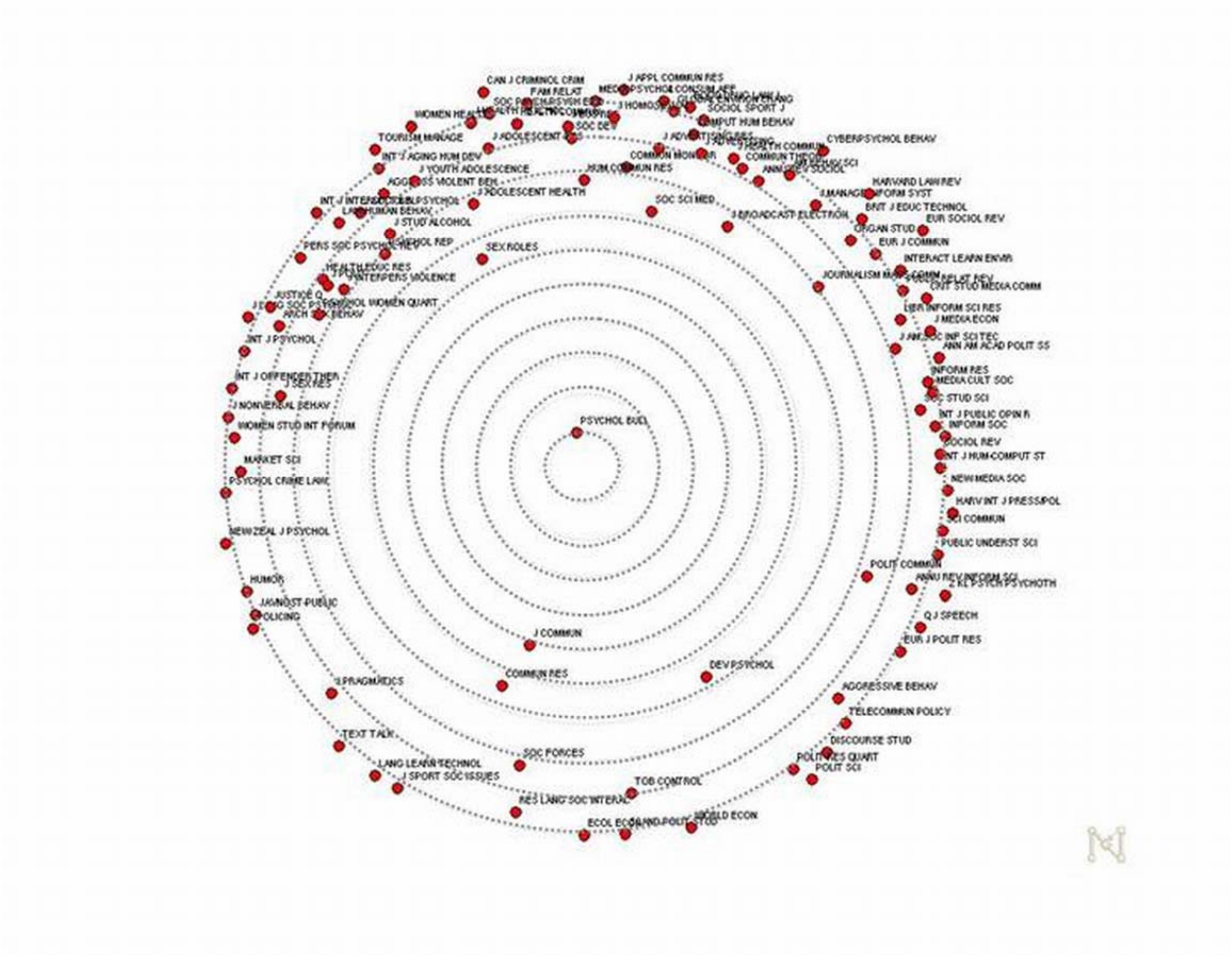

- The data used for this Figure were logarithmically scaled.

Additionally, the inter-citation matrix among 107 journals was aggregated according to the disciplinary categories listed in the ISI's database. Despite the debate about the ISI classification system (Boyack *et al.*, 2005; Leydesdorff, 2006), this perspective can be complimentary to the above findings. The 107 journals belonged to a total of 21 subject categories. The majority of journals is classified as "communication" (28 journals) or "psychology" (24 journals), followed by "Criminology & Penology" (7 journals) and "Business" (7 journals). "Information Science & Library Science", "Political Science", and "Sociology" are included with six journals. Nine categories contain only a single journal each.

According to the network analysis results (see Table 2), 24 journals belonging to "psychology" as a block were cited 2,260 times outside the psychology domain and 28 "communication" journals were cited 1,418 times by journals of other groups. The next most cited categories are "public, environmental & occupational health" (3 journals, 755 being cited) and "sociology" (6 journals, 555 being cited). Furthermore, the most preferred category by "communication" journals was "psychology" journals (565 citations). Conversely, "communication" journals were cited 487 times by "psychology" journals. This tells us that the relational strength between the two fields—"communication" and "psychology"—is the highest among all other subject pairs.

**Table 2**. Citation values for blocks of journals according to the ISI classifications

| Categories | | | | | | | | | | | | | | | | | | | | | | Sum times *citing* | Main diagonal | Nr of journals | density |
|---|---|---|---|---|---|---|---|---|---|---|---|---|---|---|---|---|---|---|---|---|---|---|---|---|---|
| BUSINESS | | 194 | 0 | 0 | 0 | 11 | 0 | 0 | 0 | 0 | 5 | 0 | 0 | 11 | 0 | 8 | 103 | 0 | 9 | 0 | 0 | 341 | 360 | 7 | 8.571 |
| COMMUNICATION | 202 | | 7 | 0 | 51 | 8 | 12 | 11 | 9 | 32 | 64 | 4 | 108 | 14 | 0 | 98 | 565 | 212 | 57 | 5 | 52 | 1511 | 2755 | 28 | 3.644 |
| CRIMINOLOGY & PENOLOGY | 0 | 51 | | 0 | 5 | 0 | 2 | 15 | 0 | 0 | 0 | 105 | 0 | 0 | 0 | 17 | 531 | 16 | 52 | 2 | 36 | 832 | 323 | 7 | 7.69 |
| ECONOMICS | 6 | 2 | 0 | | 0 | 20 | 0 | 0 | 0 | 0 | 0 | 2 | 0 | 0 | 0 | 3 | 2 | 2 | 0 | 0 | 0 | 37 | 0 | 1 | 0 |
| EDUCATION & EDUCATIONAL RESEARCH | 0 | 24 | 0 | 0 | | 0 | 4 | 0 | 0 | 0 | 0 | 0 | 0 | 0 | 0 | 0 | 127 | 92 | 3 | 0 | 0 | 250 | 5 | 4 | 0.417 |
| ENVIRONMENTAL STUDIES | 43 | 5 | 0 | 24 | 0 | | 0 | 0 | 0 | 0 | 0 | 0 | 0 | 0 | 0 | 0 | 21 | 0 | 6 | 0 | 0 | 99 | 0 | 2 | 0 |
| ERGONOMICS | 9 | 17 | 0 | 0 | 3 | 0 | | 0 | 0 | 0 | 60 | 0 | 5 | 0 | 0 | 0 | 32 | 0 | 0 | 0 | 0 | 126 | 0 | 1 | 0 |
| FAMILY STUDIES | 0 | 2 | 8 | 0 | 0 | 0 | 0 | | 3 | 0 | 0 | 2 | 0 | 0 | 0 | 2 | 47 | 3 | 13 | 3 | 0 | 83 | 0 | 1 | 0 |
| GERONTOLOGY | 9 | 22 | 0 | 0 | 0 | 0 | 0 | 11 | | 0 | 0 | 0 | 0 | 0 | 0 | 0 | 35 | 16 | 2 | 0 | 0 | 95 | 0 | 1 | 0 |
| HISTORY & PHILOSOPHY OF SCIENCE | 0 | 18 | 0 | 0 | 0 | 0 | 0 | 0 | 0 | | 0 | 0 | 0 | 0 | 0 | 0 | 2 | 0 | 6 | 0 | 0 | 26 | 0 | 1 | 0 |
| INFORMATION SCIENCE & LIBRARY SCIENCE | 32 | 144 | 0 | 0 | 2 | 0 | 38 | 2 | 0 | 23 | | 0 | 0 | 12 | 0 | 0 | 44 | 10 | 16 | 0 | 0 | 323 | 437 | 6 | 14.567 |
| LAW | 0 | 17 | 46 | 2 | 0 | 0 | 0 | 0 | 0 | 0 | 0 | | 0 | 0 | 0 | 17 | 51 | 4 | 3 | 0 | 0 | 140 | 9 | 3 | 1.5 |
| LINGUISTICS | 9 | 108 | 0 | 0 | 0 | 0 | 0 | 0 | 11 | 0 | 0 | 0 | | 0 | 0 | 0 | 51 | 2 | 0 | 6 | 0 | 187 | 10 | 2 | 5 |
| MANAGEMENT | 0 | 10 | 0 | 0 | 0 | 0 | 3 | 0 | 0 | 11 | 5 | 0 | 0 | | 0 | 2 | 9 | 4 | 31 | 0 | 0 | 75 | 0 | 1 | 0 |
| WOMEN'S STUDIES | 0 | 5 | 0 | 0 | 0 | 0 | 0 | 0 | 0 | 2 | 0 | 0 | 0 | 0 | | 0 | 30 | 7 | 5 | 0 | 0 | 49 | 0 | 1 | 0 |
| POLITICAL SCIENCE | 2 | 128 | 4 | 0 | 0 | 0 | 0 | 0 | 0 | 0 | 2 | 8 | 0 | 0 | 0 | | 23 | 3 | 35 | 0 | 0 | 205 | 224 | 6 | 7.467 |
| PSYCHOLOGY | 30 | 487 | 197 | 0 | 69 | 0 | 28 | 31 | 12 | 3 | 18 | 20 | 35 | 14 | 13 | 21 | | 352 | 184 | 15 | 162 | 1691 | 3945 | 24 | 7.147 |
| PUBLIC, ENVIRONMENTAL & OCCUPATIONAL HEALTH | 5 | 90 | 13 | 0 | 64 | 0 | 0 | 9 | 0 | 21 | 0 | 0 | 0 | 3 | 17 | 8 | 345 | | 128 | 0 | 21 | 724 | 174 | 3 | 29 |
| SOCIOLOGY | 0 | 62 | 5 | 0 | 0 | 0 | 0 | 2 | 0 | 5 | 3 | 0 | 4 | 29 | 4 | 43 | 67 | 25 | | 0 | 2 | 251 | 223 | 6 | 7.433 |
| SOCIAL SCIENCES | 0 | 23 | 0 | 0 | 0 | 0 | 0 | 0 | 0 | 0 | 0 | 0 | 4 | 2 | 0 | 2 | 98 | 0 | 2 | | 0 | 131 | 0 | 1 | 0 |
| SUBSTANCE ABUSE | 0 | 9 | 20 | 0 | 5 | 0 | 0 | 0 | 0 | 0 | 0 | 0 | 0 | 0 | 0 | 0 | 77 | 7 | 3 | 0 | | 121 | 0 | 1 | 0 |
| Sum times *cited* | 347 | 1418 | 300 | 26 | 199 | 39 | 87 | 81 | 35 | 97 | 157 | 141 | 156 | 85 | 34 | 221 | 2260 | 755 | 555 | 31 | 273 | 7297 | 8465 | 107 | |

Furthermore, this so-called (conventional) block-analysis allows the researcher to identify the extent to which the communication journals under investigation cite within their own group compared to how often they cite and are cited outside of this group. This can be done through another frequently used network metric, "density," that is, the ratio of the number of observed citations relations between journals compared to the number of possible citation relations [total cells or (N * N -1)]. The communication studies network is then considered as a valued graph. Thus, the density is calculated as the sum of the values of the observed citations between journals divided by the number of cells in the matrix. The density of "communication" journals is 3.64 (2,755 self-citations / (28 * 27 =) 756 possible citation relations) and of "psychology" journals is 7.15 (3,945 self-citations / (24 * 23 =) 552 possible citation relations). According to these density metrics, the communication network is sparser than the psychology network. The average strength of citations thus reveals that communication studies are not yet an independent inter-reading community (Van den Besselaar & Leydesdorff, 1996).

***Citation patterns among eight core journals cited by the Journal of Communication***

The analysis of citation patterns of the *Journal of Communication* and its relevant journals provides significant evidence that the authors in the seed journal, the *Journal of Communication*, do not tend to reach out to a broader literature beyond psychology. Eight journals were identified from a list of 158 journals that were cited by the *Journal of Communication* in 2006. These eight journals contribute above the threshold level of two percent of the total references in the *Journal of Communication*.

Table 3 summarizes the inter-citation frequency patterns between these journals. The left-side list of most cited journals confirms that communication scientists within the citation environment of the *Journal of Communication* are inclined to cite from journals in social and experimental psychology. Two prestigious journals in psychology, the *Personality & Social Psychology Bulletin* and *Journal of Personality & Social Psychology* were central in indegree (times cited) and outdegree values (times citing) among the journals cited by authors in the *Journal of Communication*. Note that these two journals were not included in Appendix I; articles in these journals did not cite articles in the *Journal of Communication* in 2006. Conversely, the *Psychological Bulletin* is not on this list; more specialist journals in social and experimental psychology are cited.

**Table 3**. Degree distributions of eight core journals

| Journals | Indegree (times cited by) | Journals | Outdegree (times citing to) |
|---|---|---|---|
| *PersSocPsycholB* | 1193 | *JPersSocPsychol* | 1589 |
| *JCommun* | 422 | *PersSocPsycholB* | 479 |
| *JPersSocPsychol* | 409 | *CommunRes* | 187 |
| *CommunRes* | 193 | *HumCommunRes* | 120 |
| *HealthPsychol* | 192 | *HealthPsychol* | 117 |
| *HumCommunRes* | 158 | *AmJPublicHealth* | 89 |
| *JBroadcastElectron* | 119 | *JBroadcastElectron* | 75 |
| *AmJPublicHealth* | 45 | *JCommun* | 75 |

***Citation patterns among 22 core journals citing the Journal of Communication***

To closely examine the relevant citation environment generated by the *Journal of Communication*, we selected the 22 journals that cited this seed journal, the *Journal of Communication*, in 2006, above the same two percent threshold. Twenty-two journals cite the *Journal of Communication* to the extent of more than two percent of the total number of received citations.

Table 4 clearly shows the key players in the communications field. *Communication Research* was cited 525 times followed by *Journal of Communication* (465 times). In the citing dimension, the *Communication Research* and the *Journal of Communication* cited the *Journal of Communication* 190 and 340 times, respectively. *Communication Research* cites other journals including itself less often than the *Journal of Communication*, which raises the question of what causes this journal to be cited so frequently? Although there is an effect of the smaller number of articles published in *Communication Research* (26 against 53 in the *Journal of Communication*), the main reason is that *Communication Research* is printed and distributed by a commercial publishing house (Sage), while the *Journal of Communication* is an official journal of the scholarly association (ICA). Thus, we assume that citation practices in these two journals might have been affected by different selection processes of managing editors and other diffusion channels.

Surprisingly, while the authors of the *Javnost/The Public*—a Slovenian journal published by the Faculty of Social Sciences in Ljubljana—cited 36 research papers published in journals belonging to this network, it was not cited at all by any of the 21 other journals. A leading European journal, the *European Journal of Communication*, was cited only 39 times. Furthermore, interdisciplinary journals (e.g., *Harvard International Journal of Press/Politics, New Media & Society, Annals of the American Academy of Political & Social Science, Science Communication,* and *Information Research*) were less used as sources for references than traditional communication journals.

**Table 4**. Degree distributions of core journals in the citation network of the *Journal of Communication*

| Journals | Indegree (times cited) | Journals | Outdegree (times citing) |
|---|---|---|---|
| *CommunRes* | 525 | *JCommun* | 340 |
| *JCommun* | 465 | *JournalismMassComm* | 303 |
| *JBroadcastElectron* | 285 | *CommunRes* | 190 |
| *HumCommunRes* | 281 | *HealthCommun* | 187 |
| *PolitCommun* | 202 | *JBroadcastElectron* | 177 |
| *CommunMonogr* | 179 | *JHealthCommun* | 163 |
| *JournalismMassComm* | 157 | *MediaPsychol* | 151 |
| *HealthCommun* | 97 | *AmBehavSci* | 144 |
| *JHealthCommun* | 74 | *HarvIntJPress/pol* | 137 |
| *CommunTheor* | 73 | *HumCommunRes* | 125 |
| *JAdvertising* | 72 | *CommunMonogr* | 124 |
| *IntJPublicOpinR* | 63 | *CommunTheor* | 110 |
| *SexRoles* | 62 | *IntJPublicOpinR* | 107 |
| *AmBehavSci* | 61 | *NewMediaSoc* | 104 |
| *MediaPsychol* | 54 | *JAdvertising* | 94 |
| *EurJCommun* | 39 | *SexRoles* | 79 |
| *HarvIntJPress/pol* | 30 | *PolitCommun* | 76 |
| *NewMediaSoc* | 29 | *EurJCommun* | 42 |
| *AnnAmAcadPolitSs* | 21 | *JavnostPublic* | 36 |
| *SciCommun* | 10 | *SciCommun* | 36 |
| *InformRes* | 0 | *AnnAmAcadPolitSs* | 35 |
| *JavnostPublic* | 0 | *InformRes* | 19 |

Figure 2 provides an illustration of the center-periphery structure in this network. The thickness of lines between journals at the nodes is proportional to their inter-citation frequencies. In other words, the width of the lines between journals can be considered as the communication strength among authors who publish their research in the particular journals. The arrow head means the direction of citation. The majority of journals in the network are strongly connected to the journals (e.g., *Communication Research, Journal of Communication*) that would fall into the categories of American scholarship and social-psychology oriented research traditions.

However, European and interdisciplinary journals tend to appear at the margins on both sides of the network diagram. For example, the *Journal of Advertising* is a preferred journal among communication scientists as well as business scholars majoring in marketing. *Political Communication* is a joint publication of the ICA and the American Political Science Association. *Science Communication* publishes research related to public understanding of science that is one of the key areas in science study.

**Figure 2**. Network diagram of 22 journals in the core of the citation network of the *Journal of Communication.*

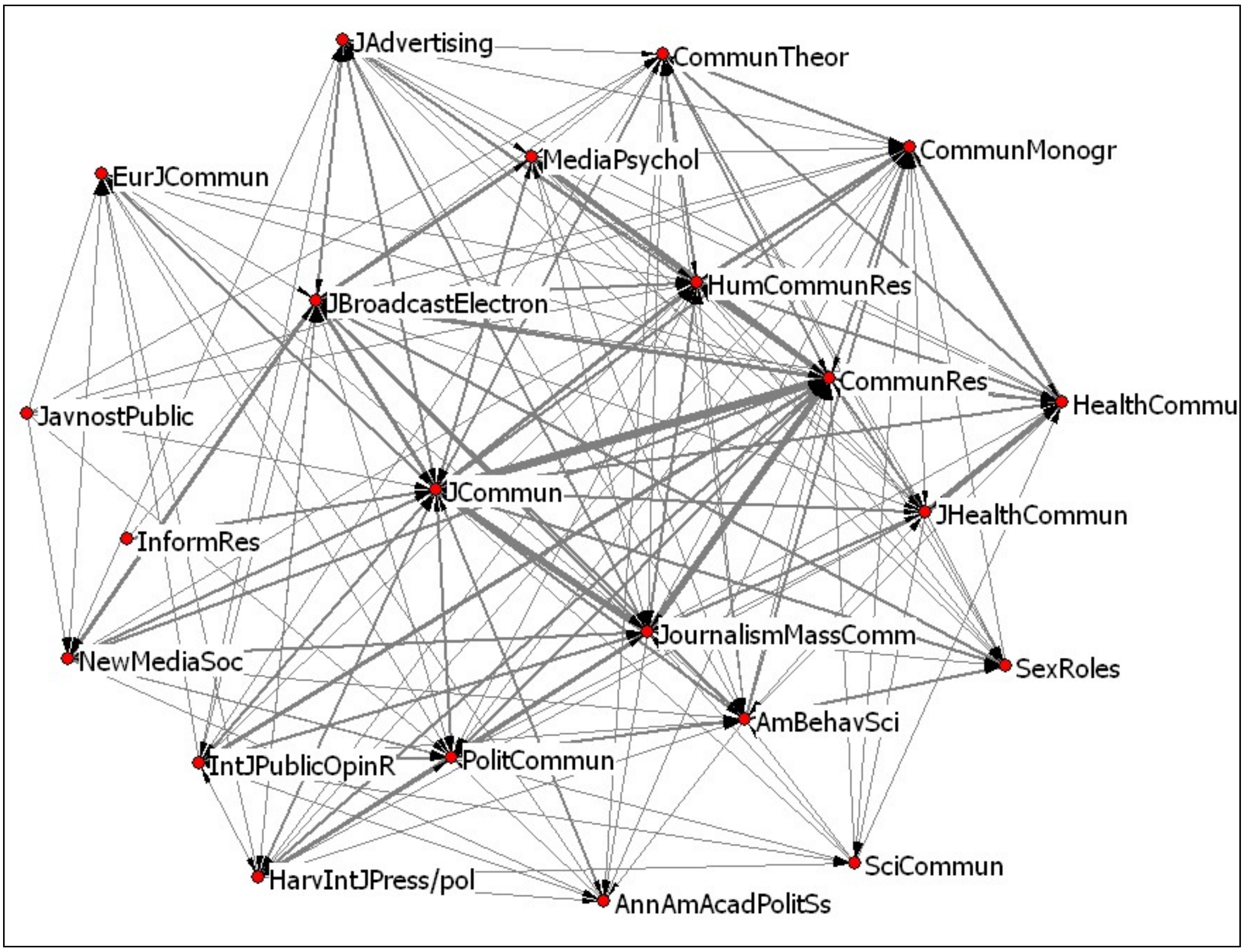

**Table 5**. The citation matrix of 22 journals in the core of the citation network of the *Journal of Communication.*

| | | | | | | | | | | | | | | | | | | | | | | |
|---|---|---|---|---|---|---|---|---|---|---|---|---|---|---|---|---|---|---|---|---|---|---|
| AmBehavSci | | 3 | 13 | 11 | 2 | | 6 | 3 | 6 | | 2 | 5 | 16 | 15 | 6 | | 10 | | 7 | 26 | | 13 |
| AnnAmAcadPolitSs | | | | 4 | 2 | 5 | | | | | 8 | | | 12 | | | | | | 4 | | |
| CommunMonogr | | | | 23 | 11 | | 2 | 6 | 32 | | | | 10 | 26 | 2 | | | 6 | | 3 | 3 | |
| CommunRes | | | 29 | | 5 | 3 | 3 | 4 | 55 | | 3 | | 25 | 18 | | | 11 | 5 | | 19 | | 10 |
| CommunTheor | | 2 | 15 | 27 | | | | | 21 | | 2 | 2 | 11 | 13 | 4 | | 3 | 3 | | | | 7 |
| EurJCommun | | | | 8 | | | | | 2 | | | | 6 | 12 | | | 4 | | 2 | 8 | | |
| HarvIntJPress/pol | 3 | 4 | | 20 | | 3 | | | 7 | | 2 | 4 | 10 | 27 | | | 14 | | | 41 | 2 | |
| HealthCommun | 7 | | 40 | 15 | 13 | | | | 25 | | | 7 | 10 | 26 | 37 | | 7 | | | | | |
| HumCommunRes | 3 | | 30 | 51 | 5 | | | 5 | | | | 2 | 5 | 12 | 2 | | 4 | | | 4 | | 2 |
| InformRes | | | | | | | | | | | | | 7 | 12 | | | | | | | | |
| IntJPublicOpinR | | 4 | | 31 | | 4 | 2 | | 5 | | | 2 | 3 | 27 | | | 14 | | | 15 | | |
| JAdvertising | 2 | | 4 | 15 | 3 | | | | 8 | | 5 | | 18 | 20 | | | 5 | 5 | 2 | | | 7 |
| JBroadcastElectron | 2 | | 4 | 36 | 4 | 2 | 4 | | 21 | | 7 | 4 | | 45 | 5 | | 22 | 7 | 3 | 11 | | |
| JCommun | 6 | 2 | 18 | 93 | 13 | 5 | 2 | 12 | 37 | | 12 | 14 | 45 | | 11 | | 16 | 18 | 3 | 29 | | 4 |
| JHealthCommun | 2 | | 6 | 20 | 8 | | 2 | 59 | 11 | | | 5 | 5 | 18 | | | 13 | | | 2 | 3 | 9 |
| JavnostPublic | | | 2 | | 3 | 4 | | | 7 | | | | | 11 | | | | | 5 | 4 | | |
| JournalismMassComm | 6 | 4 | 7 | 71 | | 6 | 7 | 2 | 16 | | 14 | 16 | 29 | 68 | 5 | | | 5 | 7 | 34 | | 6 |
| MediaPsychol | 4 | | 3 | 48 | 4 | | | 4 | 17 | | | 4 | 33 | 26 | 2 | | 2 | | | | | 4 |
| NewMediaSoc | 9 | 2 | | 15 | | 3 | | | 6 | | | 4 | 32 | 15 | | | 14 | | | 2 | 2 | |
| PolitCommun | 2 | | 3 | 16 | | 4 | 2 | | 2 | | 4 | | 4 | 28 | | | 11 | | | | | |
| SciCommun | 2 | | 2 | 10 | | | | 2 | | | 4 | | | 11 | | | 5 | | | | | |
| SexRoles | 13 | | 3 | 11 | | | | | 3 | | | 3 | 16 | 23 | | | 2 | 5 | | | | |

***Subgroup structures among the 107 journals citing the Journal of Communication in 2006***

In the above section, the individual position of each journal in the network was detected in terms of citation relations and then the structure among the core journals was identified using a traditional threshold routine. This section employs techniques from social network analysis that can be used to examine the sub-structures of a network, based on both the direct and indirect citation relations among journals. "Component analysis" enables the researcher to identify where an entire network is divided (Hanneman & Riddle, 2005).

The network of 107 journals under investigation is one component because the journals were selected on the criterion that they cited the *Journal of Communication.* In this case, component analysis for valued network can be used to examine the hierarchy among the components. The resulting dendrogram can be used to separate components. A hierarchical dendrogram, or a tree diagram, can be used to illustrate the arrangement of the sub-groups. The distinction between components is then made according to a cut-off value. In this research, a cut-off value of 22 was assumed because sub-groups clearly emerged at this level. Due to space limitation, the dendrogram is not printed here and only the resulting components are described.

The overall structure of the network appeared to have one large component with relatively tight couplings among journals and several minor components. Special attention is placed on the largest component and its sub-component parts because all major communication journals belong to this component. Forty-two journals are connected within this component. This major component can be again dissected into three primary sectors: general psychology, health-related psychology, and communication research. This finding confirms that the underlying structure of communication study is strongly integrated with social and experimental psychology.

**Table 4**. Three primary groups in the largest component

| Groups | Journals |
|---|---|
| General psychology | *AggressiveBehav, JAdolescentHealth, PsycholBull, TobControl DevPsychol, JAdolescentRes, JHealthPsychol, JSocClinPsychol, JYouthAdolescence, PersSocPsycholRev, PsycholRep, PsycholWomenQuart, SexRoles, SocDev, SocPsychPsychEpid, SocSciMed, WomenHealth* |
| Health-related psychology | *JStudAlcohol, AggressViolentBeh, AnnuRevSociol, ArchSexBehav, CommunMonogr, HealthCommun, HealthEducRes, JHealthCommun, JHomosexual, JInterpersViolence, JSexRes, SocForces* |
| Communication research | *AmBehavSci, CommunRes, CommunTheor, HarvIntJPress/pol, HumCommunRes, IntJOffenderTher, IntJPublicOpinR, JBroadcastElectron, JCommun, JournalismMassComm, MediaPsychol, NewMediaSoc, PolitCommun* |

In the remaining component hierarchy, interesting smaller units were found. For example, five marketing communication journals (*Journal of Advertising, Journal of Advertising Research, Journal of Business Research, Marketing Science, Tourism Management*) constitute one sub-group. Information science journals also form a single group: *Annual Review of Information Science & Technology, Journal of American Society for Information Science & Technology, Information Research, Library & Information Science Research, International Journal of Human-Computer Studies, Journal of Management Information Systems, Computers in Human Behavior, and Cyberpsychology & Behavior*. There is another group of political science journals (*Journal of Polity, Political Research Quarterly*) and a legal studies-related component (*Law & Human Behavior, Psychology Crime & Law, Organization Studies, Sociological Review*).

Six journals are isolates, that is, not connected to any other journals in the network. Four of these isolates (*Food & Drug Law Journal, Language Learning & Technology, Political Science, and Text & Talk*) did not receive any citations from any other journal in the network. Additionally, “bi-component” analysis was used to find the so-called, key weak node in a network (Hanneman & Riddle, 2005). If the *Journal of Communication* were removed, *Language Learning & Technology* would become completely isolated in the network. Compared to the other three isolates, *Language Learning & Technology* did not have any citation in either the cited or citing dimensions.

**Discussion**

In this research, we found that the impact of social and experimental psychology journals on communication studies is so large that they can be considered as a major source of theories and methods for the authors in communication journals. Communication researchers have tended to employ social psychology literature as a reference frame. This accords with Roger's (1994) statement in his book entitled "A History of Communication Study" that the major intellectual force behind the emergence of communication studies is social psychology (p. 491). The dominant perspective of communication scholarship is empirical and quantitative with a strong focus on determining the effects of communication. The dependent character of communication studies on social and experimental psychology was clearly shown by this citation analysis at the level of journals.

The underlying structure identified from the component analysis gives strong support to the notion that communication studies are overwhelmingly influenced by literature from social and experimental psychology, and to a much smaller extent by literature in political science. We found one large cluster made up of social psychology and mainstream communication journals. However, network analysis enabled us to discern small clusters that have emerged as independent specialties within this context. In the case of advertising research, its intellectual background also comes from theories and methods in social psychology, but it is currently institutionalized as a stand-alone school in a number of universities. The University of Illinois at Urbana-Champaign established the Department of Advertising in 1949. This was the first such academic department in the America. However, several universities, for example, the University of Texas at Austin and Michigan State University at East Lansing, have started recently to treat advertising as another scholarly unit. New media studies seem to follow similar paths. We suggest that this movement toward diversification within communication studies may result in a further fragmentation in the structural patterns of relations among social psychology and communication journals. However, the findings of this research did not reflect an incremental development of knowledge at the core of the field of communication studies itself.

One can argue that this could be a consequence of choosing the *Journal of Communication* as a seed journal. What if we used another journal as our focal node? Might we have found more political science in the background? The answer is yes and no. As voiced in a preliminary analysis based on the 2004 ISI data (Leydesdorff & Park, 2006), the field of communication studies is insufficiently codified in theoretical terms to stand alone in terms of its intellectual traditions and resources. In a genealogical network, children having the same biological parents have biological affinity. In a similar way, scientists who were trained and influenced by the same advisors with the authors of articles in the *Journal of Communication* are more likely to adopt and share knowledge in psychology and political science. Probably communication scholarship is not mature enough to split into more coherent sub-networks that appear to be integrated among themselves. It seems unlikely that any other intellectual paradigm will replace the dominant psychology tradition in the near future.

Journals which can be considered as European in terms of their publishing houses and affiliations of members of the editorial board had weak network connectivity with other journals when compared to US-affiliated journals. For example, the *Javnost/The Public* cited 36 research papers published in the 107 journals belonging to the network drawn from the *Journal of Communication*, but it was not cited at all by the other journals. According to Lauf's (2005, p. 146) national diversity scores in articles published between 1998 and 2002 in the ISI set of communication journals, the *Javnost/The Public* was the next most internationalized (.95 out of 1.00), after *Discourse & Society* (.96). Another leading European journal, the *European Journal of Communication,* occupied the 9th position (.80) on Lauf's ranking list, but we found that it was a relatively infrequently cited journal. The address of the publishing and editorial houses may not tell us much about the character of a journal, because this market is internationalized. Nonetheless, as Lauf (2005) points out, geographical location can be expected to help deepen the intensity of linkage structures among journals in neighboring regions and provide prospective authors with better accessibility for subscription, submission, and citation.

The examination of "self-citations within a journal" provided another interesting conclusion. While the cited and citing numbers show inter-journal traffic, these self-citations are the number of citations made to the articles published in the journal from articles in the same journal. Thus, this can reveal the specific journal's internal cohesion among current and prospective authors. The self-citations in European journals only loosely linked with influential US communication journals are extremely low. For instance, *Javnost/The Public* and the *European Journal of Communication* had two and thirteen self-citations, respectively. This means that communication scholars publishing in these journals seem to perceive themselves as marginal sources to acquire relevant information for their research.

Overall, the application of this citation-based, research design has not provided more finely grained and qualitative information about structural relationships within communication studies. In a well-written and concise overview, Contractor (1996, pp. 110-112) classified communication studies into two approaches: one focusing on the various situational contexts in which communication may occur (e.g., interpersonal communication, organization communication, mass communication, cross-cultural communication, health communication, and technologically mediated communication), and the other on various functions of communication including persuasion, socialization, and conflict resolution. Whatever categories communication scientists label themselves, their theoretical, methodological and empirical backgrounds are largely attributable to the literature in social psychology. As Contractor emphasized, as the study of communication moves into the twenty-first century, three major paradigmatic alternatives (the *systems perspective,* based on network analysis among a social system's components; the *interpretive perspective,* underscoring the meaning of messages uncovered by the receiver; and the *critical perspective,* concerned with social reality in relation to the power relation) will slowly emerge in addition to the earlier psychology-dominated approach. However, this now more than ten-year old vision remains speculative without data that show these emerging developments in communication studies.

**Acknowledgements**

An earlier version of this paper was presented to the conference of the KSJCS (Korean Society of Journalism and Communication Studies), 24-26 April 2008, Jeju Island, South Korea. http://www.ksjcs.or.kr/english/english_ksjcs.html . This research was supported by YeungNam University research grants in 2008.

Appendix 1. Names and citation rates for the 107 journals in the citation impact environment of the *Journal of Communication* in 2006.

| Name of journal | Journal Abbreviation (ISI) | Nr of papers in 2006 | Cited references | Total cites | Within-journal "self-citations" | Journal Impact Factor | Country |
|---|---|---|---|---|---|---|---|
| *Aggression and Violent Behavior* | AggressViolentBeh | 44 | 440 | 171 | 42 | 1.600 | England |
| *Aggressive Behavior* | AggressiveBehav | 55 | 424 | 289 | 129 | 1.012 | USA |
| *American Behavioral Scientist* | AmBehavSci | 87 | 376 | 243 | 57 | 0.466 | USA |
| *Annals of the American Academy of Political and Social Science* | AnnAmAcadPolitSs | 86 | 94 | 111 | 32 | 0.736 | USA |
| *Annual Review of Information Science and Technology* | AnnuRevInformSci | 13 | 209 | 151 | 76 | 1.385 | USA |
| *Annual Review of Sociology* | AnnuRevSociol | 19 | 130 | 352 | 44 | 3.275 | USA |
| *Archives of Sexual Behavior* | ArchSexBehav | 55 | 470 | 377 | 182 | 2.198 | USA |
| *British Journal of Educational Technology* | BritJEducTechnol | 61 | 63 | 55 | 48 | 0.406 | England |
| *Canadian Journal of Criminology and Criminal Justice* | CanJCriminolCrim | 36 | 91 | 81 | 53 | 0.367 | Canada |
| *Communication Monographs* | CommunMonogr | 22 | 235 | 309 | 39 | 0.909 | England |
| *Communication Research* | CommunRes | 26 | 332 | 727 | 53 | 1.056 | USA |
| *Communication Theory* | CommunTheor | 21 | 192 | 132 | 26 | 1.050 | USA |
| *Computers in Human Behavior* | ComputHumBehav | 70 | 271 | 245 | 164 | 0.808 | USA |
| *Critical Studies in Media Communication* | CritStudMediaComm | 23 | 50 | 45 | 18 | 0.475 | USA |
| *Cyberpsychology & Behavior* | CyberpsycholBehav | 85 | 316 | 218 | 142 | 1.061 | USA |
| *Developmental Psychology* | DevPsychol | 107 | 757 | 1264 | 416 | 3.556 | USA |
| *Discourse Studies* | DiscourseStud | 40 | 87 | 41 | 15 | 0.471 | England |
| *Ecological Economics* | EcolEcon | 244 | 842 | 831 | 805 | 1.223 | Netherlands |
| *European Journal of Communication* | EurJCommun | 22 | 64 | 84 | 13 | 0.429 | England |
| *European Journal of Political Research* | EurJPolitRes | 43 | 187 | 191 | 141 | 1.916 | Netherlands |
| *European Sociological Review* | EurSociolRev | 36 | 100 | 51 | 36 | 0.607 | England |
| *Family Relations* | FamRelat | 51 | 150 | 148 | 67 | 0.731 | USA |
| *Food and Drug Law Journal* | FoodDrugLawJ | 30 | 52 | 46 | 46 | 0.397 | USA |
| *Global Environmental Change-Human and Policy Dimensions* | GlobalEnvironChang | 30 | 147 | 140 | 120 | 2.600 | England |
| *Harvard International Journal of Press-Politics* | HarvIntJPress/pol | 19 | 188 | 67 | 30 | 0.525 | USA |
| *Harvard Law Review* | HarvardLawRev | 42 | 420 | 414 | 393 | 7.863 | USA |
| *Health Communication* | HealthCommun | 53 | 417 | 216 | 73 | 1.169 | USA |
| *Health Education Research* | HealthEducRes | 90 | 335 | 310 | 118 | 1.623 | England |
| *Human Communication Research* | HumCommunRes | 21 | 246 | 439 | 53 | 1.372 | England |
| *Humor-International Journal of Humor Research* | Humor | 15 | 105 | 73 | 62 | 0.421 | England |
| *Information Research-An International Electronic Journal* | InformRes | 44 | 209 | 71 | 36 | 0.870 | England |
| *Information Society* | InformSoc | 29 | 116 | 78 | 36 | 0.803 | USA |
| *Interactive Learning Environments* | InteractLearnEnvir | 17 | 25 | 9 | 4 | 0.300 | England |
| *International Journal of Aging & Human Development* | IntJAgingHumDev | 34 | 140 | 80 | 45 | 0.614 | USA |
| *International Journal of Human-Computer Studies* | IntJHumComputSt | 81 | 256 | 217 | 130 | 1.094 | England |
| *International Journal of Intercultural Relations* | IntJInterculRel | 45 | 264 | 164 | 133 | 0.578 | USA |
| *International Journal of Offender Therapy and Comparative Criminology* | IntJOffenderTher | 45 | 170 | 96 | 36 | 0.750 | USA |
| *International Journal of Psychology* | IntJPsychol | 58 | 82 | 92 | 26 | 0.571 | France |
| *International Journal of Public Opinion Research* | IntJPublicOpinR | 27 | 164 | 120 | 43 | 0.522 | England |
| *Javnost-the Public* | JavnostPublic | 18 | 51 | 15 | 2 | 0.051 | Slovenia |
| *Journal of Adolescent Health* | JAdolescentHealth | 241 | 844 | 798 | 481 | 2.710 | USA |
| *Journal of Adolescent Research* | JAdolescentRes | 24 | 139 | 222 | 26 | 1.582 | USA |
| *Journal of Advertising* | JAdvertising | 39 | 440 | 388 | 176 | 0.667 | USA |
| *Journal of Advertising Research* | JAdvertisingRes | 40 | 277 | 412 | 178 | 0.478 | USA |
| *Journal of Applied Communication Research* | JApplCommunRes | 18 | 114 | 63 | 19 | 0.719 | England |
| *Journal of Broadcasting & Electronic Media* | JBroadcastElectron | 29 | 345 | 465 | 114 | 0.704 | USA |
| *Journal of Business Research* | JBusRes | 151 | 394 | 302 | 215 | 0.815 | USA |
| *Journal of Communication* | JCommun | 53 | 603 | 872 | 91 | 1.159 | USA |
| *Journal of Consumer Affairs* | JConsumAff | 16 | 71 | 61 | 27 | 0.718 | USA |
| *Journal of Health Communication* | JHealthCommun | 62 | 521 | 266 | 138 | 1.387 | USA |
| *Journal of Health Psychology* | JHealthPsychol | 65 | 373 | 192 | 108 | 1.267 | England |
| *Journal of Homosexuality* | JHomosexual | 78 | 445 | 269 | 173 | 0.233 | USA |
| *Journal of Interpersonal Violence* | JInterpersViolence | 92 | 459 | 486 | 182 | 1.139 | USA |
| *Journal of Language and Social Psychology* | JLangSocPsychol | 21 | 156 | 77 | 28 | 0.821 | USA |
| *Journal of Management Information Systems* | JManageInformSyst | 41 | 343 | 322 | 240 | 1.818 | USA |
| *Journal of Media Economics* | JMediaEcon | 12 | 39 | 43 | 17 | 0.125 | USA |
| *Journal of Nonverbal Behavior* | JNonverbalBehav | 13 | 95 | 91 | 39 | 1.240 | USA |
| *Journal of Politics* | JPolit | 71 | 237 | 385 | 141 | 1.055 | USA |
| *Journal of Pragmatics* | JPragmatics | 85 | 288 | 336 | 219 | 0.465 | Netherlands |
| *Journal of Sex Research* | JSexRes | 33 | 269 | 433 | 100 | 1.417 | USA |
| *Journal of Social and Clinical Psychology* | JSocClinPsychol | 55 | 244 | 223 | 53 | 1.471 | USA |
| *Journal of Sport & Social Issues* | JSportSocIssues | 21 | 65 | 73 | 24 | 0.675 | USA |
| *Journal of Studies on Alcohol* | JStudAlcohol | 108 | 599 | 751 | 478 | 1.884 | USA |
| *Journal of the American Society for Information Science and Technology* | JAmSocInfSciTec | 142 | 721 | 825 | 544 | 1.555 | USA |
| *Journal of Youth and Adolescence* | JYouthAdolescence | 86 | 692 | 410 | 150 | 1.214 | USA |
| *Journalism & Mass Communication Quarterly* | JournalismMassComm | 33 | 490 | 322 | 104 | 0.688 | USA |
| *Justice Quarterly* | JusticeQ | 22 | 138 | 126 | 57 | 1.038 | USA |
| *Language Learning & Technology* | LangLearnTechnol | 12 | 51 | 49 | 49 | 1.310 | USA |
| *Law and Human Behavior* | LawHumanBehav | 40 | 290 | 303 | 174 | 2.122 | USA |
| *Library & Information Science Research* | LibrInformSciRes | 28 | 128 | 113 | 34 | 1.059 | USA |
| *Marketing Science* | MarketSci | 38 | 656 | 700 | 621 | 3.977 | USA |
| *Media Culture & Society* | MediaCultSoc | 40 | 106 | 119 | 37 | 0.418 | USA |
| *Media Psychology* | MediaPsychol | 20 | 218 | 86 | 20 | 1.152 | USA |
| *New Media & Society* | NewMediaSoc | 46 | 188 | 90 | 31 | 0.988 | England |
| *New Zealand Journal of Psychology* | NewZealJPsychol | 18 | 55 | 21 | 12 | 0.325 | USA |
| *Organization Studies* | OrganStud | 77 | 281 | 291 | 206 | 1.583 | England |
| *Personality and Social Psychology Review* | PersSocPsycholRev | 20 | 167 | 123 | 19 | 3.348 | USA |
| *Policing-An International Journal of Police Strategies & Management* | Policing | 40 | 90 | 40 | 37 | 0.200 | USA |
| *Political Communication* | PolitCommun | 24 | 139 | 301 | 39 | 1.118 | England |
| *Political Research Quarterly* | PolitResQuart | 52 | 218 | 95 | 37 | 0.468 | USA |
| *Political Science* | PolitSci | 4 | 29 | 7 | 7 | 0.269 | New Zealand |
| *Psychological Bulletin* | PsycholBull | 37 | 432 | 1689 | 165 | 12.725 | USA |
| *Psychological Reports* | PsycholRep | 267 | 447 | 625 | 237 | 0.364 | USA |
| *Psychology Crime & Law* | PsycholCrimeLaw | 46 | 227 | 83 | 53 | 1.015 | England |
| *Psychology of Women Quarterly* | PsycholWomenQuart | 37 | 380 | 496 | 134 | 1.096 | USA |
| *Public Relations Review* | PublicRelatRev | 71 | 189 | 183 | 126 | 0.296 | USA |
| *Public Understanding of Science* | PublicUnderstSci | 23 | 125 | 107 | 59 | 0.978 | England |
| *Quarterly Journal of Speech* | QJSpeech | 17 | 122 | 145 | 97 | 0.333 | England |
| *Research On Language and Social Interaction* | ResLangSocInterac | 13 | 60 | 112 | 24 | 1.000 | USA |
| *Scandinavian Political Studies* | ScandPolitStud | 20 | 47 | 39 | 25 | 0.342 | England |
| *Science Communication* | SciCommun | 17 | 90 | 60 | 18 | 0.800 | USA |
| *Sex Roles* | SexRoles | 152 | 1336 | 1075 | 523 | 0.942 | USA |
| *Social Development* | SocDev | 38 | 320 | 160 | 54 | 1.349 | England |
| *Social Forces* | SocForces | 86 | 316 | 448 | 164 | 1.214 | USA |
| *Social Psychiatry and Psychiatric Epidemiology* | SocPsychPsychEpid | 133 | 353 | 297 | 194 | 1.577 | Germany |
| *Social Science & Medicine* | SocSciMed | 530 | 2296 | 2284 | 1661 | 2.749 | England |
| *Social Studies of Science* | SocStudSci | 33 | 121 | 192 | 95 | 1.426 | USA |
| *Sociological Review* | SociolRev | 50 | 117 | 105 | 41 | 0.705 | England |
| *Sociology of Sport Journal* | SociolSportJ | 19 | 115 | 118 | 60 | 0.773 | USA |
| *Telecommunications Policy* | TelecommunPolicy | 36 | 128 | 134 | 110 | 0.705 | England |
| *Text & Talk* | TextTalk | 27 | 65 | 4 | 4 | 0.000 | Germany |
| *Tobacco Control* | TobControl | 95 | 552 | 680 | 460 | 2.797 | England |
| *Tourism Management* | TourismManage | 113 | 444 | 391 | 372 | 0.856 | England |
| *Women & Health* | WomenHealth | 46 | 225 | 140 | 54 | 0.815 | USA |
| *Womens Studies International Forum* | WomenStudIntForum | 52 | 79 | 64 | 30 | 0.462 | England |
| *World Economy* | WorldEcon | 81 | 58 | 45 | 41 | 0.655 | Netherlands |
| *Zeitschrift fur Klinische Psychologie und Psychotherapie* | ZKlPsychPsychoth | 28 | 114 | 52 | 50 | 0.517 | Germany |